\documentclass[12pt]{article}
\usepackage{epsf}
\usepackage{amsmath}

\usepackage{graphics}
\usepackage{cite}

\begin{document}

\hfill January 2026

\begin{center}

{\bf \large Response to  2512.23460[hep-ph] by Czarnecki and Khanna}\\
{\bf ("False vacuum decay and flaws in Frampton's model for the origin of life")}.

\vspace{1.00cm}

{\bf P. H. Frampton}\footnote{paul.h.frampton@gmail.com}\\
\vspace{0.5cm}
{\it Dipartimento di Matematica e Fisica "Ennio De Giorgi",\\ 
Universit\`{a} del Salento and INFN-Lecce,\\ Via Arnesano, 73100 Lecce, Italy.
}

\vspace{1.0in}
\end{center}

\begin{abstract}
\noindent
We respond to a critique 2512.23460[hep-ph] of our paper 2505.05634[hep-ph].
\end{abstract}

\vspace{1.5cm}
\begin{center}
{\bf Text}
\end{center}

\bigskip

\noindent
The most substantial criticism by the authors of 2512.23460[hep-ph] 
concerning \\2505.05634[hep-ph] is the claim that
we inserted Planck's constant into our calculation based only on dimensional
arguments. This is incorrect as Planck's constant appears naturally in the
quantum tunnelling amplitude. More complete formulas can be found in
our ref[8] which was, for some reason, not cited by Czarnecki and Khanna, but
was the first paper in which the correct result for vacuum decay 
in relativistic quantum field theory appeared. Although
2505.05634[hep-ph] made a number of assumptions, this was not one of them.
We reaffirm that, given the
assumptions we did make, it is exceedingly likely that humankind will remain isolated and lonely forever.

\end{document}